\documentclass[aps,  prl, reprint, amssymb, twocolumns, shortbibliography]{revtex4-2}

\usepackage{verbatim}
\usepackage[english]{babel}
\usepackage{hyperref} 
\usepackage{svg}
\usepackage{graphicx}
\usepackage{dcolumn}
\usepackage{bm}
\usepackage{amsmath}
\usepackage{enumerate}
\usepackage[normalem]{ulem}
\usepackage{float}
\usepackage{refcount}

\usepackage{enumitem}

\newcommand{\be}{\begin{equation}}
\newcommand{\ee}{\end{equation}}
\newcommand{\bae}{\begin{eqnarray}}
\newcommand{\eae}{\end{eqnarray}}

\begin{document}

\newtheorem{deff}{Definition}
\newtheorem{con}{Construction}
\newtheorem{note}{Note}
\newtheorem{remark}{Remark}
\newtheorem{casei}{Case}
\newtheorem{assumption}{Assumption}

\newtheorem{thm}{Theorem}
\newtheorem{lem}[thm]{Lemma}
\newtheorem{cor}[thm]{Corollary}
\newtheorem{prop}[thm]{Proposition}

\newtheorem{ex}{Example}[section]
\newtheorem{convention}{Convention}[section]


\title{Driven criticality links universal computation and optimal representations}

\author{Adrián Roig}
\affiliation{Departamento de Electromagnetismo y F{\'\i}sica de la Materia and Instituto Carlos I
de F{\'\i}sica Te{\'o}rica y Computacional. Universidad de Granada.E-18071, Granada, Spain}

\author{Miguel A. Mu\~noz$^\dagger$} 
\affiliation{Departamento de Electromagnetismo y F{\'\i}sica de la Materia and Instituto Carlos I
de F{\'\i}sica Te{\'o}rica y Computacional. Universidad de Granada.E-18071, Granada, Spain}

\author{Guillermo B. Morales$^\dagger$}
\affiliation{Departamento de Electromagnetismo y F{\'\i}sica de la Materia and Instituto Carlos I
de F{\'\i}sica Te{\'o}rica y Computacional. Universidad de Granada.E-18071, Granada, Spain}

\date{\today}

\begin{abstract}
Near-critical dynamics are often linked to enhanced computation, but the underlying mechanism remains unclear. We address this question in reservoir computing, where a fixed recurrent network maps input sequences into high-dimensional states and only a simple readout is trained. We extend fixed-reservoir universality results to discrete-time input-driven reservoirs and connect their key geometric condition, neighborhood separation, to dynamics. To this end, we introduce a finite-resolution neighborhood separability index and an input-conditioned maximal Lyapunov exponent. We find that neighborhood separability, chaotic time-series prediction, and smooth high-dimensional representation geometry are optimized in the same narrow window of marginal driven stability. In this regime, the covariance spectrum approaches the power-law scaling expected for near-optimal smooth representations. Our results link edge-of-instability computation, universality, readout performance, and optimal representation geometry within a common dynamical framework.
\end{abstract}

\maketitle

\vspace{-0.4em}

\begin{center}
\small
$^\dagger$ These authors contributed equally.
\end{center}

\begin{figure}[tbh!]
    \centering
    \includegraphics[width=1.0\linewidth]{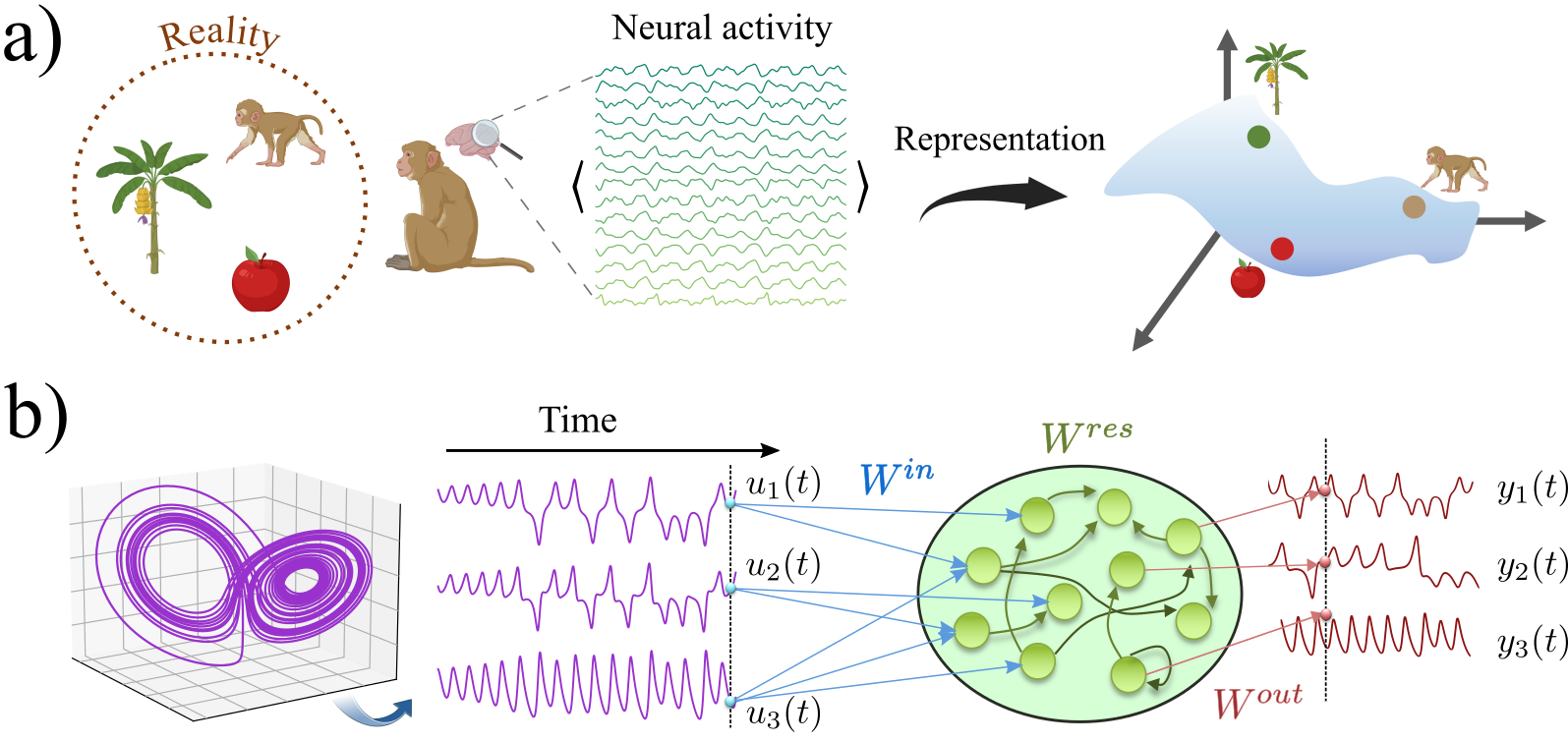}
  \caption{\textbf{Representation map and reservoir-computing architecture.}
(a) External stimuli are encoded as neural activity patterns forming a trajectory or point cloud in high-dimensional state space. 
(b)  A multivariate input stream, $ u(t)$, drives a fixed recurrent network through $W^{\rm in}$, producing reservoir states, $ x(t)$. Only the readout matrix, $W^{\rm out}$, is trained to produce the target, $ y(t)$.}
\label{fig:1}
\end{figure}

A recurring theme in complex recurrent systems is that operating near the boundary between distinct dynamical regimes or phases can yield an unusual combination of sensitivity, dynamic range across many scales, and flexible yet robust responses; features often linked to enhanced computational capacity \cite{Langton1990, Packard1988, Crutchfield1988, Mitchell1993, BerNat2004, Mora-Bialek, RMP}. In neuroscience, this idea is captured by the criticality hypothesis: cortical circuits may operate close to a critical regime to support efficient information transmission, memory, and rapid adaptation \cite{BP2003, Chialvo2010, Plenz-functional, Shew-review, Breakspear-review, Priesemann-review, Beggs2022, Obyrne-review}. 
Empirically, near-critical brain dynamics have been probed through scale-invariant activity, long-range correlations, and correlation-based signatures of proximity to criticality \cite{BP2003,Palva2014,Meshulam2019,Dahmen_second_2019, Hu-Sompo, Morales2023}.
A closely related theme appears in machine learning, particularly in the reservoir computing (RC) paradigm \cite{Schrauwen2007,Seoane2019}, where a fixed recurrent neural network (the reservoir) maps an input stream into a high-dimensional dynamical internal state, while only a simple readout layer is trained to perform a task (see 
Fig.~\ref{fig:1}b)
\cite{Jaeger2001,Jaeger2002,Jaeger_harnessing_2004,Jaeger2007,Maass2002,maass_liquid_2011,lukosevicius_practical_2012,BerNat2004,Boedecker2011,Legenstein_edge_2007, Prokopenko2011, Morales-optimal, singh2025contraction, Dambre2012}. 
Since pioneering studies of computation at the edge of chaos, such systems have often been reported to perform best near the boundary between strongly stable dynamics and unstable, perturbation-amplifying regimes
\cite{Boedecker2011,Legenstein_edge_2007, Prokopenko2011, Morales-optimal}.

A particularly relevant advance has been the formulation of functional optimality in terms of representation geometry \cite{Stringer2019,Nassar_1n_2020,Dahmen_second_2019,Morales-optimal}. Here, a \emph{representation} is the map from an external input, stimulus, or task parameter to the corresponding network state, viewed as a point or trajectory in a high-dimensional state space (see Fig.~\ref{fig:1}a). Stringer \emph{et al.} \cite{Stringer2019} proposed that near-optimal representations should be as high-dimensional as possible while remaining smooth, and showed that for a smooth representation of a $d_\ell$-dimensional input manifold the covariance eigenspectrum must decay as $\lambda_n\sim n^{-\mu}$, with $\mu\geq\mu_c=1+2/d_\ell$, so that the broadest possible smooth representation lies near $\mu\simeq\mu_c$. 

Since scale-free spectra are a hallmark of critical phenomena \cite{Binney, Hu-Sompo, Morales2023}, this observation raises the question of whether the dynamical regimes that produce high-dimensional optimal representations arise at the boundary between stable and unstable dynamics. While this possibility has been explored computationally by some of us \cite{Morales-optimal}, a principled explanation of why such a regime should support both optimal representations and computational power remains lacking.

Here we take a step in this direction by building on recent universality results of Sugiura and collaborators \cite{Sugiura2024existence,Sugiura2024-non,Sugiura2025necessary}.
 Universality denotes the property that a fixed reservoir, by training only a polynomial readout, can approximate \emph{any} target dynamical input--output map to arbitrary accuracy. In Sugiura's \emph{et al.} framework, universality is guaranteed by a local necessary and sufficient condition known as the \emph{neighborhood separation property} (NSP): non-overlapping input neighborhoods must remain distinguishable after they are represented in reservoir state space. In contrast to earlier approaches based on strong fading-memory assumptions \cite{Grigoryeva2018}, NSP is a local separability condition on input neighborhoods.

We explore these ideas in a minimal reservoir-computing model driven by chaotic time series, using the input-conditioned maximal Lyapunov exponent (MLE) to track the reservoir's driven stability as recurrent connectivity is varied. We find that NSP and task performance peak sharply in a narrow marginally stable regime, $\mathrm{MLE}\simeq0^-$, which is also where the covariance-spectrum exponent most closely approaches the optimal smoothness bound of Stringer \emph{et al.}

Thus, our central goal is to connect the reservoir's dynamical regime to the geometry of its representation. We argue that marginal driven stability is the regime where two geometric requirements can coexist: local neighborhood separation, needed for universal computation, and smooth high-dimensional structure, needed for near-optimal representations. In this view, edge-of-instability performance reflects a dynamical balance between sensitivity and regularity.

Let us consider a standard reservoir-computing setup in which a fixed recurrent network of $N$ nodes transforms an input stream into a high-dimensional state, and only a readout is trained \cite{Jaeger_harnessing_2004,Jaeger2001,lukosevicius_practical_2012} (see Fig.~\ref{fig:1}b). The reservoir evolves in discrete time steps according to the input-driven  dynamics:
\begin{equation}
 x_t=\phi\left(\rho W^{res}  x_{t-1}+\varepsilon W^{in}  u_{t-1}\right),
\label{eq:Model}
\end{equation}
where $ x_t\in\mathcal M\subset\mathbb R^N$ is the reservoir state at time $t$, $ u_{t-1}\in\mathcal U\subset\mathbb R^d$ is the input at the previous time step, with $\mathcal U$ bounded and $d\ll N$, and $\phi$ is a smooth component-wise activation function. We use $\phi(y)=\tanh(y)$ as a representative bounded sigmoidal nonlinearity (with general assumptions stated in SI, Sec.~S1.2). For $\phi=\tanh$, the reservoir state space is the N-dimensional hypercube
$\mathcal M=[-1,1]^N$.
The recurrent matrix, $W^{res}\in\mathbb R^{N\times N}$, and input matrix, $W^{in}\in\mathbb R^{N\times d}$, are drawn with i.i.d.\ Gaussian entries of zero mean and variances $1/N$ and $1/d$, respectively; and the two fundamental hyperparameters of the model are $\rho$, which controls the effective recurrent gain; and $\varepsilon$, which sets the input strength, so the autonomous limit is recovered for $\varepsilon=0$.

For any finite input sequence \(u_{0:T-1}\equiv (u_0,\ldots,u_{T-1})\),
Eq.~\eqref{eq:Model} maps the sequence into the final reservoir state,
which we call its representation:
\begin{equation}
x_T=\mathbf f(u_{0:T-1}) .
\label{eq:fmap}
\end{equation}
The set of all finite sequences with entries in \(\mathcal U\) is denoted
by \(\mathbb V^*\), so \(\mathbf f:\mathbb V^*\to\mathbb R^N\).

Our aim is not to identify yet another performance ``sweet spot,'' but
to determine which dynamical regimes support \emph{universality}: the
ability of a fixed reservoir, with only the readout trained, to
approximate arbitrary continuous tasks of the input sequence (End
Matter, Sec.~B.1). To make this precise, we equip \(\mathbb V^*\) with a fading-memory metric, so that it becomes an input-sequence space in which sequences are close when they differ little in recent inputs, while differences in the remote past are weighted less strongly (End Matter, Sec.~A.1).

In this geometry, the neighborhood separation property (NSP) requires
distinct local neighborhoods of sequences to remain separated after
mapping into reservoir state space by \(\mathbf f\). Equivalently, NSP
constrains the local non-degeneracy of the inverse relation
\(\mathbf f^{-1}\): nearby reservoir states should not ambiguously
correspond to unrelated input sequences (End Matter, Sec.~A.2).

Sugiura and collaborators proved that this condition is necessary and sufficient for universal approximation with polynomial readouts in input-driven continuous-time reservoirs~\cite{Sugiura2024existence,Sugiura2024-non,Sugiura2025necessary}. Here, we have extended these universality results to \emph{discrete-time} reservoirs (End Matter, Sec.~A.1, and SI Secs.~S1.1 and S1.2), showing that the same NSP criterion guarantees universality with polynomial readouts in our setting. Thus, NSP and the Stringer \emph{et al.} smoothness bound enter a common geometric language: the Stringer bound constrains the regularity of \(\mathbf f\), whereas NSP constrains the local non-degeneracy of \(\mathbf f^{-1}\).

\begin{figure}[tbh!]
    \centering
    \includegraphics[width=\linewidth]{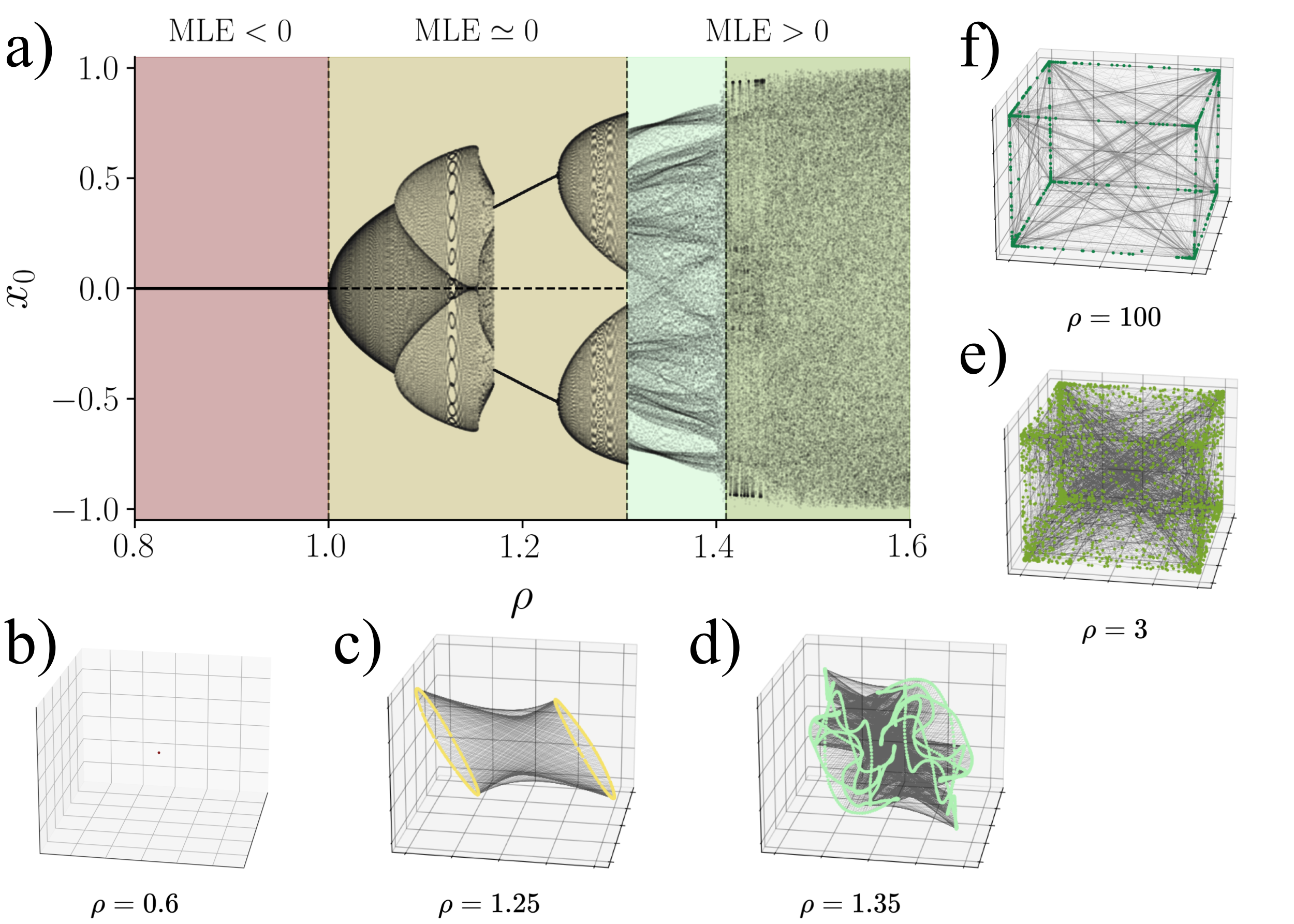}
\caption{\textbf{Autonomous reservoir dynamics changes with recurrent gain.}
(a) Unforced dynamics ($\varepsilon=0$, $N=200$) as the recurrent gain $\rho$ is increased. The panel shows the state $x_0$ of a representative reservoir unit versus $\rho$, with the typical structure of a bifurcation diagram. Shaded regions indicate the corresponding driven dynamical regimes classified by the input-conditioned maximal Lyapunov exponent (MLE): contracting dynamics ($\mathrm{MLE}<0$), marginal driven stability ($\mathrm{MLE}\simeq0$), and expanding dynamics ($\mathrm{MLE}>0$).
(b--f) Representative reservoir trajectories in state space for increasing $\rho$: weak recurrence collapses the representation near a fixed point (b); near marginal stability, trajectories unfold into a structured geometry (c); beyond stability, stretching and folding lead to mixing (d); and at very large $\rho$, saturation drives activity onto the faces, edges, and vertices of the state-space hypercube (e,f). Gray lines connect consecutive reservoir states.}
\label{fig:2}
\end{figure}

To examine how different dynamical regimes affect NSP, we vary the gain parameter \(\rho\), which controls the relative strength of recurrence and therefore how input sequences are embedded into reservoir states, \(x_T=\mathbf f(u_{0:T-1})\). We first use the isolated reservoir, \(\varepsilon=0\), to identify the underlying autonomous dynamical regimes, and then interpret the input-driven case, \(\varepsilon>0\), as a perturbation of these regimes. In the autonomous limit, increasing \(\rho\) produces a classical bifurcation scenario (Fig.~\ref{fig:2}): reservoir units move from a globally attracting fixed point to oscillatory dynamics and chaos, before saturation drives activity toward the boundaries (faces, edges, vertices) of the state-space hypercube. Although the transition values depend on the nonlinearity and reservoir realization, this phenomenology is generic for bounded sigmoidal activation functions.

In the weakly recurrent regime (left of Fig.~\ref{fig:2}, \(\rho\ll1\)), the driven reservoir cannot propagate past information far into the future. Indeed, in the limiting case \(\rho=0\), one has \(x_t=\phi(\varepsilon W^{\rm in}u_{t-1})\) for \(\varepsilon\neq0\), so the map \(\mathbf f\) collapses all input sequences sharing the same last value into the same representation. Consequently, $\mathbf f$ is not injective: no inverse function $\mathbf f^{-1}$ exists on its image that can recover the input sequence uniquely from the representation, precluding NSP. 
More generally, a Taylor expansion of Eq.~\eqref{eq:Model}
in \(\rho\), truncated at order \(n\), shows that \(x_T\) depends only
on the most recent \(O(n)\) inputs; to that order, sequences that
coincide on their last \(n\) entries map to the same state,  
only a short effective memory is available and NSP does not hold in practice.
A different route to NSP failure occurs as $\rho$ increases into expanding, mixing regimes: nearby inputs may separate rapidly, but bounded dynamics stretch and fold their images, so nearby reservoir states can originate from unrelated input sequences and local inverse discriminability is lost. Saturation provides a limiting form of this bounded-state-space distortion. In the large-$\rho$ limit (Fig.~\ref{fig:2}, right), the autonomous dynamics is driven toward the boundaries of the state space $\mathcal M$, namely the hypercube (SI Secs.~S1.3 and S1.4). In saturated directions, small input perturbations are filtered by the local Jacobian, i.e., to order $\varepsilon$,
\begin{equation}
\phi(\rho W^{\rm res} x+\varepsilon W^{\rm in} u)
\simeq
\phi(\rho W^{\rm res} x)
+
\varepsilon J_\phi(\rho W^{\rm res} x)W^{\rm in} u,
\end{equation}
so that, for $\phi=\tanh$, the diagonal entries of $J_\phi$ are $1-\tanh^2[(\rho W^{\rm res}x)]$ and vanish in saturated units. Thus, the directions that dominate the state become locally insensitive to the input, violating the local non-degeneracy required by NSP.

Together, weak recurrence and mixing-induced folding, which culminates in saturation, obstruct NSP in complementary ways. Increasing $\rho$ toward marginal stability makes higher-order recurrent effects dynamically relevant and extends memory; beyond this window, however, uncontrolled folding destroys the smooth geometry of the input representation (SI Secs.~S1.3 and S1.4). Marginal driven stability is therefore the natural candidate regime in which memory, local separation, and a smooth, optimal input-representation geometry can coexist.

To examine this semi-analytical picture computationally, we use a representative reservoir realization with fixed architecture ($W^{\rm res}$, $W^{\rm in}$, $\phi=\tanh$), small input gain $\varepsilon$, and varying $\rho$, as a numerical instance of the generic mechanism described above. Inputs are normalized time series, $ u_t\in\mathcal U\subset\mathbb R^3$, generated from the Lorenz attractor \cite{Lorenz1963,Strogatz_nonlinear_2000} (Fig.~\ref{fig:1}b). Starting from an arbitrary point $ u_0\in\mathbb R^3$ on the attractor, we construct length-$T$ sequences:
\begin{equation}
u_{0:T-1}=\bigl( u_0,\,g( u_0),\,\ldots,\,g^{T-1}( u_0)\bigr),
\end{equation}
where $g:\mathbb R^3\to\mathbb R^3$ denotes a discrete-time flow map of the Lorenz dynamics. The terminal state \(x_T=\mathbf f(u_{0:T-1})\) should retain task-relevant information to be used by the readout.

The remaining question is which dynamical regime makes the induced representation $\mathbf f$ both locally separable and useful for readout-based approximation. To connect the underlying dynamical regime with NSP, we introduce two measures. First, we characterize the stability of the driven reservoir dynamics through an input-conditioned maximal Lyapunov exponent (MLE), which measures the rate of amplification of infinitesimal state perturbations along trajectories driven by a given input sequence (SI Sec.~S2.2) \cite{Hart2024ReservoirCLE,Uchida2008LocalConditional,Pyragas1997Conditional,Pecora1991Driving}. 
Second, we introduce the neighborhood separability index (NSI), a new finite-resolution diagnostic that quantifies how closely a sampled reservoir representation realizes the neighborhood separation required by NSP (SI Sec.~S2.1).
 For each pair of nearby but disjoint input neighborhoods in the fading-memory metric, we map the corresponding inputs through the reservoir and compare the separation between the resulting state clouds with their internal spread. Specifically, if $X_A$ and $X_B$ are the two image clouds, with empirical centroids $m(X_A),m(X_B)$ and radii $r(X_A),r(X_B)$, NSI is based on the local margin:
\begin{equation}
\Delta_{AB}=\|m(X_A)-m(X_B)\|_2-\bigl[r(X_A)+r(X_B)\bigr],
\end{equation}
where $\|\cdot\|_2$ denotes the Euclidean norm in reservoir state space. Positive margins certify finite-resolution neighborhood separation, whereas negative margins indicate overlap of the corresponding envelope of image clouds. NSI is obtained by averaging the positive part of this margin over sampled neighborhood pairs. 
Thus, NSI is a geometric diagnostic: a computable, finite-resolution
counterpart of NSP for readout-based interpolation (SI Sec.~S2.1 and
Figs.~S3--S5).
 \begin{figure}[tbh!]
    \centering
\includegraphics[width=0.9\linewidth]{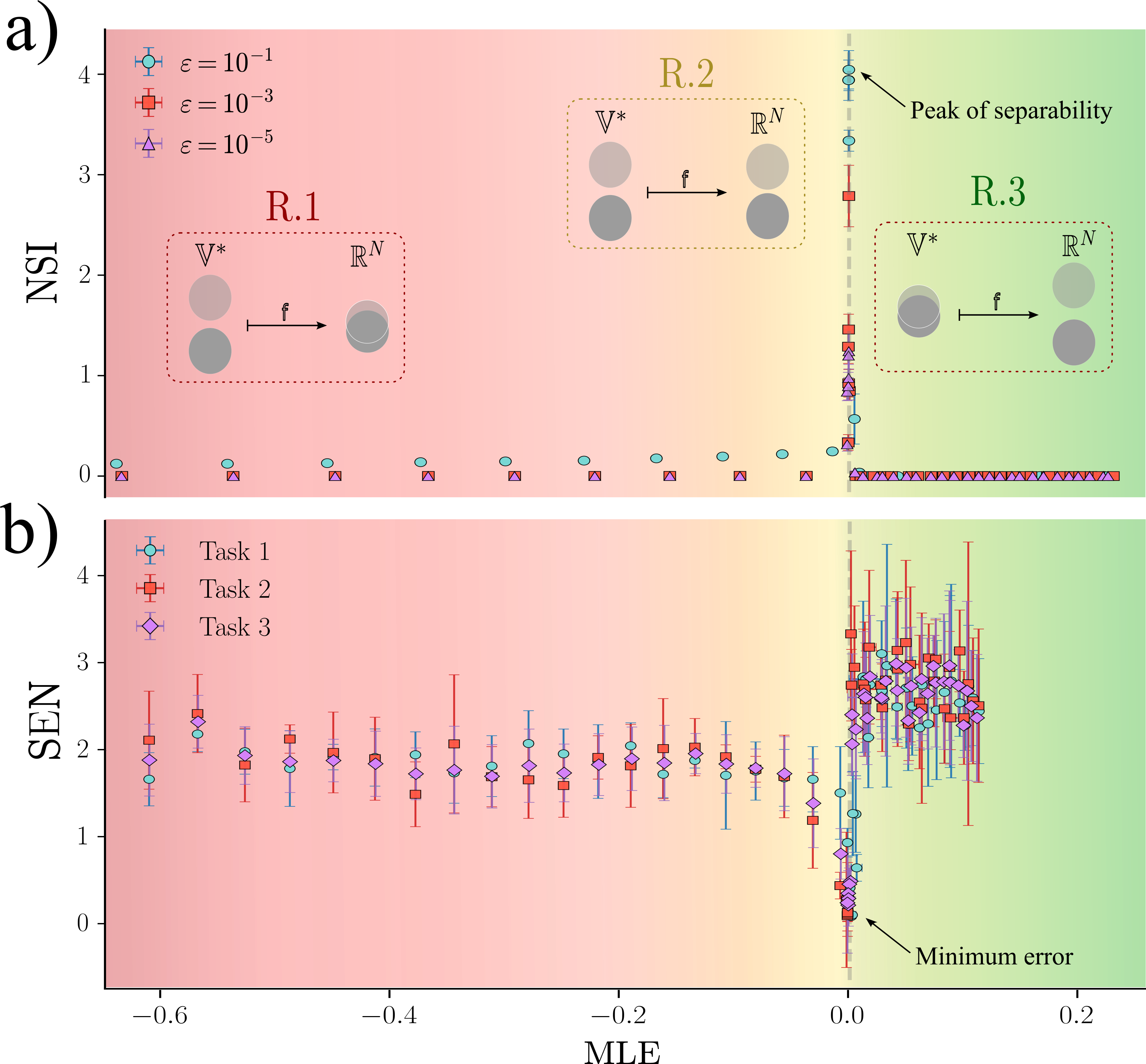}
\caption{\textbf{Neighborhood separability and performance peak near
marginal driven stability.}
(a) Neighborhood separability index (NSI) versus input-conditioned
maximal Lyapunov exponent (MLE), obtained by varying \(\rho\) for
\(N=2000\) (SI Secs.~S2.1 and S2.2). Error bars indicate variability
across sampled neighborhood pairs. Shaded regions indicate the regimes
identified in Fig.~\ref{fig:2}: strongly contractive R1, marginally
stable R2, and unstable/saturation-dominated R3. NSI is maximal near
\(\mathrm{MLE}\simeq0^{-}\).
(b) Supremum-norm error (SEN) for benchmark tasks (End Matter,
Sec.~B.2, and SI Sec.~S2.3), with \(N=2000\) and \(\varepsilon=0.01\).
SEN is minimized in the same R2 window.}
\label{fig:3}
\end{figure}

\begin{figure*}[tbh!]
    \centering
\includegraphics[width=\linewidth]{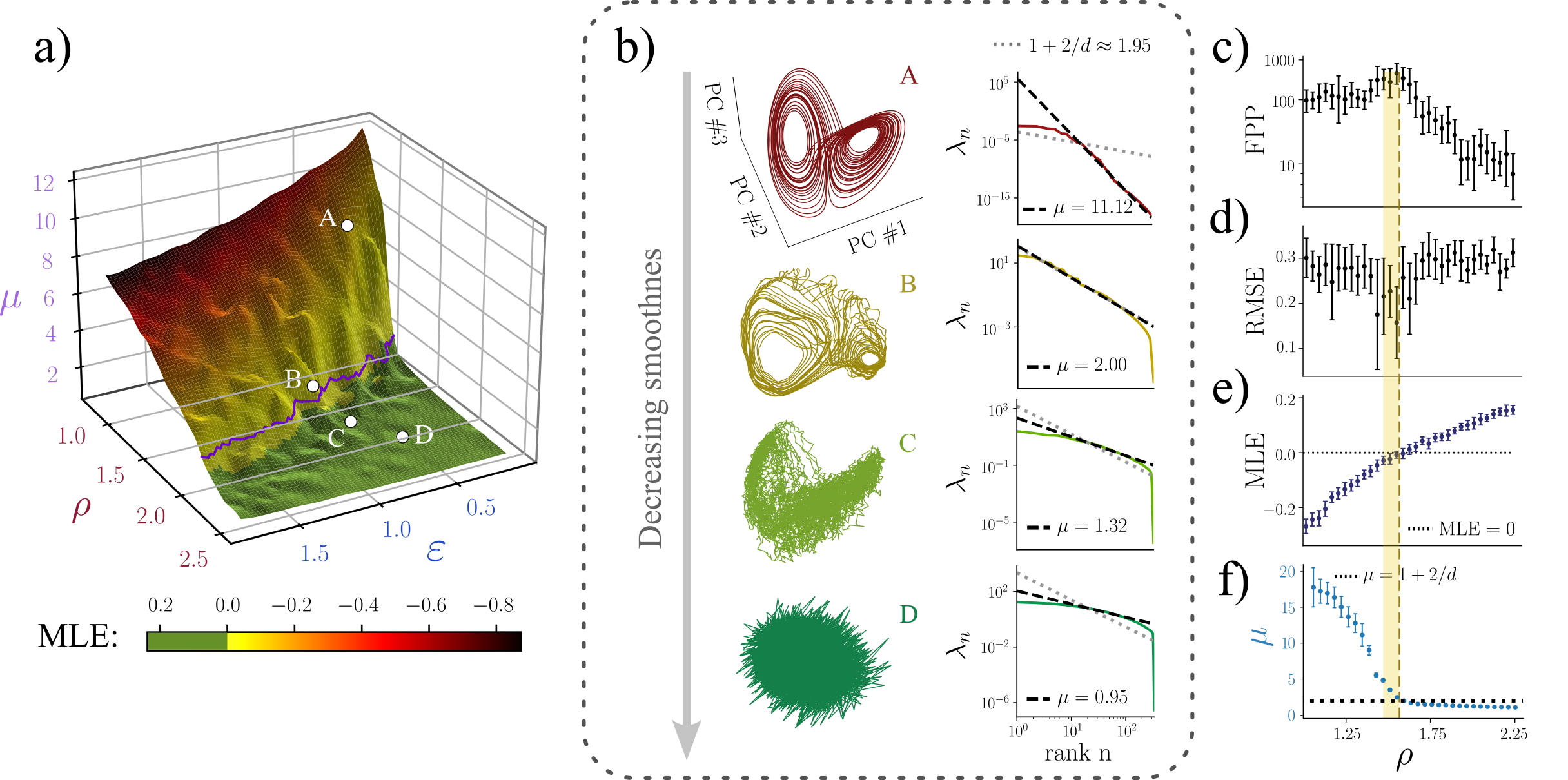}
\caption{\textbf{Marginal driven stability meets optimal representation geometry and predictive power.}
(a) Input-conditioned maximal Lyapunov exponent (MLE) as a function of spectral radius $\rho$ and input scale $\varepsilon$.
(b) PCA projections of reservoir activity for four representative parameter choices, together with the corresponding covariance spectra.
(c, d) Furthest-predicted point (FPP) and Root-mean-square error (RMSE), showing optimal prediction near marginal driven stability, $\mathrm{MLE}\simeq 0^{-}$. 
(e, f)  Input-conditioned MLE and spectral exponent $\mu$ as a function of $\rho$, showing $\mu\approx 1+2/d_{\ell}$ near the point of marginal driven stability, $\mathrm{MLE}\simeq 0^{-}$.}
\label{fig:4}
\end{figure*}

Figure~\ref{fig:3}a shows NSI as a function of the input-conditioned MLE as $\rho$ is varied, revealing three regimes. At small $\rho$, region R1 is strongly contractive, with $\mathrm{MLE}<0$: input sequences are rapidly forgotten, nearby input neighborhoods overlap in reservoir state space, and NSI is negligible. 
At intermediate $\rho$, region R2 approaches marginal driven stability,
$\mathrm{MLE}\simeq0^{-}$, where robustly positive NSI indicates
finite-resolution separation and locally coherent geometry in $\mathcal M$.
At larger $\rho$, region R3 is unstable and/or saturation-dominated, and NSI collapses again because input geometry is degraded by suppressed sensitivity or uncontrolled stretching and folding. The same R2 window minimizes the test supremum-norm error (SEN; Fig.~\ref{fig:3}b and End Matter, Sec.~B.2) for all benchmark tasks, showing that neighborhood separability and readout performance are optimized together near marginal driven stability. Although NSP guarantees universality for polynomial readouts, Fig.~\ref{fig:3} further indicates that near this optimum a simple linear readout already suffices for the tasks considered here.

The coincidence of separability and optimal performance in Fig.~\ref{fig:3} suggests that the marginally stable regime should also be distinguished by the geometry of its internal representations. To quantify the linear degrees of freedom available to the readout (SI Secs.~S1.5 and S1.7), we analyze the covariance matrix of reservoir states sampled along driven trajectories:
\begin{equation}
C=
\mathbb E_{u\in \mathbb V^*}
\left[x(u)x(u)^\top\right],
\qquad
x(u)=\mathbf f(u)\in\mathbb R^N,
\end{equation}
together with its rank-ordered eigenvalues $\lambda_1\ge\cdots\ge\lambda_N$. 
Over an intermediate range of parameters, the spectrum follows a power law, $\lambda_n\sim n^{-\mu}$, with the fitted exponent $\mu$ varying across the $(\rho,\varepsilon)$ plane, as shown in Fig.~\ref{fig:4}a.  Smaller $\mu$ corresponds to slower spectral decay and therefore to a broader representation, with more active linear modes available to the readout.

Fig.~\ref{fig:4}b shows reservoir trajectories projected onto the leading principal components for four representative regimes. For small $\rho$, the representation closely mimics the Lorenz attractor and deterministically tracks the input, but task performance remains low in this low-dimensional representation regime. As $\rho$ increases at fixed $\varepsilon$, the trajectories spread across more principal directions, reaching their largest effective dimensionality near the stability boundary. Comparing the fitted spectral exponent with the Stringer {\emph et al.} smoothness bound, $\mu_c=1+2/d_\ell$, shows that this regime approaches the highest-dimensional representation compatible with the bound. For the Lorenz attractor, the intrinsic dimension gives $\mu_c\simeq1.95$, in agreement with the exponent measured where NSI is positive, as shown in Fig.~\ref{fig:4}f.

Thus, the optimal regime is not merely high-dimensional; it also respects the smoothness constraint set by the dimensionality of the input space. We have therefore connected two geometric requirements: the Stringer \emph{et al.} bound constrains the forward regularity of $\mathbf f$, whereas NSP constrains the local non-degeneracy of the inverse relation $\mathbf f^{-1}$ (see the \textit{covering-learning protocol}, SI Sec.~S1.6). Near marginal stability, both requirements are approximately satisfied: the reservoir-state manifold remains smooth enough to approach $\mu_c$ while preserving finite-resolution neighborhood separation.

In more unstable or saturation-dominated regimes, forward smoothness and/or local inverse discriminability deteriorate, the usable linear degrees of freedom become less reliable, and predictive performance worsens (Fig.~\ref{fig:4}c,d; see End Matter, Sec.~B.2 for the furthest-predicted point, FPP, and root-mean-square error, RMSE). Taken together, Figs.~\ref{fig:3} and \ref{fig:4} show that the best-performing regime combines input-dependent separation with a broad, high-dimensional but regular linear encoding, explaining why a simple linear readout suffices in our benchmarks.

In conclusion, we identify a dynamical mechanism underlying edge-of-instability optimality in input-driven reservoirs. Marginal driven stability balances memory, sensitivity, and regularity: input sequences are neither erased by strong contraction nor degraded by uncontrolled amplification, folding, or saturation. This balance allows the reservoir representation to approach the smooth high-dimensional geometry predicted by the Stringer \emph{et al.} bound while retaining the finite-resolution neighborhood separation required for universal computation.

Our results bring together three complementary perspectives. First, by extending Refs.~\cite{Sugiura2024existence,Sugiura2024-non,Sugiura2025necessary} to discrete time, we show that NSP remains a necessary and sufficient condition for universality with polynomial readouts. Second, NSI and the input-conditioned MLE provide an operational bridge from this theory to finite reservoirs: as $\rho$ is varied, finite-resolution neighborhood separability becomes robust and task performance is optimized near marginal driven stability, $\mathrm{MLE}\simeq0^{-}$. Third, this marginally stable regime also yields broad covariance spectra approaching the Stringer \emph{et al.} bound, $\mu\simeq1+2/d_\ell$, thereby linking computational performance to smooth high-dimensional representation geometry. In this way, our results connect edge-of-instability computation, universality theory, and the geometry of near-optimal representations within a common dynamical framework.

Several open avenues remain for future work, including extending these results to continuous-time reservoirs and identifying self-organization mechanisms that tune reservoirs toward marginal driven stability for broad classes of inputs and tasks. Such mechanisms could make neighborhood separability, predictive performance, and smooth high-dimensional representations robust without external fine tuning, providing design principles for optimized physical reservoirs across different substrates \citep{Tanaka2019}.

\medskip
{\bf{Acknowledgments:}} 
We acknowledge the Spanish Ministry
and Agencia Estatal de Investigación (AEI), MICIN/AEI/10.13039/501100011033, for financial support, Project PID2023-149174NB-I00 funded also by ERDF/EU. 
We acknowledge inspiring discussions with V. Buendía and R. Calvo.
\bibliography{Bibliography}
\clearpage

\newpage
\section*{End Matter}

\subsection{Appendix A: Fading-memory sequence space }
\subsubsection{A.1. Adapting neighborhood-separation universality to the discrete-time framework}

We consider finite input sequences with entries in a compact set
$\mathcal U\subset \mathbb{R}^d$. For each $T\geq 1$, let
\begin{equation}
    \mathbb V^*_{T}=\{(v_{-T+1},\ldots,v_0):v_{-k}\in \mathcal U\},
\end{equation}
and define the set of finite sequences as $\mathbb V^*=\bigcup_{T\geq 1}\mathbb V^*_T$.
The use of negative indices is only notational: it identifies $v_0$
as the most recent input and $v_{-k}$ as an input $k$ steps in the past.

To formulate neighborhood separation, $\mathbb V^*$ is embedded into a compact
metric space. Let $\mathbb V=\mathcal U^{\mathbb Z_-}$ be the space
of left-infinite input sequences. In the Supplemental Material we construct an
extended fading-memory metric $\widetilde d$ on
$\widetilde{\mathbb V}=\mathbb V\cup\mathbb V^*$ using a decreasing weight sequence
for past coordinates together with a length penalty for finite sequences (SI, Sec. S1.1). This
metric has two key properties:
\begin{equation}
    (\widetilde{\mathbb V},\widetilde d )\ \text{is compact},
    \ \ 
    \mathbb V^*\ \text{is dense in}\ \widetilde{\mathbb V}.
\end{equation}
In particular, if $v=(\ldots,v_{-2},v_{-1},v_0)\in\mathbb V$, its finite
truncations $v^{[T]}=(v_{-T+1},\ldots,v_0)\in\mathbb V^*$ converge to $v$ in
$(\widetilde{\mathbb V},\widetilde d)$ as $T\to\infty$.

This compactification is used to define neighborhoods of finite sequences and to
state NSP in the same compact-metric framework as in Refs.~\cite{Sugiura2025necessary,Sugiura2024existence, Sugiura2024-non}. Importantly,
the reservoir representation $\mathbf f:\mathbb V^*\to\mathbb R^N$ must be bounded and is not assumed
to be continuous on $\widetilde{\mathbb V}$. The role of NSP is to avoid imposing a strong global
fading-memory condition on the reservoir map, replacing it by a local separation
requirement: distinct neighborhoods in the compactified input space must
have separated images under $\mathbf f$ (SI, Sec. S1.2, S1.2.1).

\subsubsection{A.2. Neighborhood separability property}
In the compactified space \((\widetilde {\mathbb V },\widetilde d )\), the
reservoir map is defined only on the dense subset of finite sequences,
\begin{equation}
    \mathbf f:\mathbb V^*\to \mathbb R^N .
    \label{A2.1}
\end{equation}
Accordingly, neighborhoods are taken in the compactified space and then
restricted to finite sequences.

We say that \(\mathbf f\) satisfies the
neighborhood separation property (NSP) on \(\mathbb V^*\) if, for every pair of
distinct finite sequences \(u,v\in \widetilde{ \mathbb V}\), there exists \(\delta>0\) such that:
\begin{equation}
    \overline{\mathbf f(N_\delta(u))}\cap \overline{\mathbf f(N_\delta(v))}=\varnothing,
     \label{A2.2}
\end{equation}
where
\begin{equation}
    N_\delta(x)=B_\delta(x)\cap \mathbb V^*, 
    \quad 
    B_\delta(x)=\{y\in \widetilde{\mathbb V}:\widetilde d(x,y)<\delta\}.
     \label{A2.3}
\end{equation}
Equivalently, distinct compactified input sequences possess neighborhoods whose finite-sequence traces are mapped by the reservoir into disjoint subsets of state space.

\subsection{Appendix B: Universality, readout learning, and task performance}

\subsubsection{B.1. Universality for uniform approximations}

Let \(h_{\mathrm{target}}:\mathbb V^*\to\mathbb R^d\) denote the target task. We assume that \(h_{\mathrm{target}}\) is the restriction to the finite-sequence space \(\mathbb V^*\) of a continuous function defined on the compactified input space \(\widetilde{\mathbb V}\). Given \(\varepsilon>0\), we say that the reservoir map \(\mathbf f:\mathbb V^*\to\mathbb R^N\) is universal for uniform approximation if there exists a polynomial readout \(p:\mathbb R^N\to\mathbb R^d\) such that:
\begin{equation}
\sup_{u\in\mathbb V^*}
\left\|
h_{\mathrm{target}}(u)-(p\circ \mathbf f)(u)
\right\|<\epsilon .
\end{equation}
For scalar tasks this reduces componentwise to the corresponding scalar inequality.
We refer to the associated uniform error as the supremum-norm error (SEN).

\subsubsection{B.2. Benchmark tasks}

Theorem-level universality is formulated for polynomial readouts, whereas the numerical benchmarks in the paper use linear readouts for computational economy.
The three tasks analyzed in Fig.~\ref{fig:3} are one-step prediction, past reconstruction, and a fading-memory average.
For an input sequence $u=(u_{-T+1},\ldots,u_{0})$, with $g$ denoting the discrete-time Lorenz flow map, the corresponding target maps are
\begin{align}
 &h^{(1)}_{\rm target}(u_{-T+1},\ldots,u_{0})
=
g(u_{0}), \\
& h^{(2)}_{k,\rm target}(u_{-T+1},\ldots,u_{0})
=
u_{-k},
\ 0\leq k\leq T-1 ,
\end{align}
whenever $k<T$ and equal to a fixed value if $k\geq T$ (SI, Sec. S2.3),
and
\begin{equation}
h^{(3)}_{\rm target}(u_0,\ldots,u_{T-1})
=
\frac{1}{T}\sum_{n=0}^{T-1}\beta^{T-1-n}u_n ,
\end{equation}
for some $1>\beta>0$. For each sampled sequence $u$, let $x_u=\mathbf f(u)\in\mathbb R^N$ denote the final reservoir state and let $h_u=h_{\rm target}(u)\in\mathbb R^d$ denote the corresponding target.
We fit a linear readout $\pi:\mathbb R^N\to\mathbb R^d$, $\pi(x)=A x$, by ridge regression on the training set $\mathbb V^*_{\rm train}\subset \mathbb V^*$:
\begin{equation}
A^*
=
\operatorname*{arg\,min}_{A\in\mathbb R^{d\times N}}\left\{
\sum_{u\in\mathbb V^*_{\rm train}}
\|A x_u-h_u\|^2
+
\lambda \|A\|_F^2\right\},
\end{equation}
for a small ridge regression coefficient $\lambda>0$, and $\|\cdot\|_F$ is the Frobenius norm.
Performance is then evaluated on a disjoint test set $\mathbb V^*_{\rm test}\subset \mathbb V^*$ using the SEN,
\begin{equation}
\mathrm{SEN}
=
\sup_{u\in\mathbb V^*_{\rm test}}
\left\|
A^*\mathbf f(u)-h_{\rm target}(u)
\right\| .
\end{equation}
Thus, the benchmarks test whether the reservoir representation is sufficiently rich and regular for low-complexity readouts to approximate the target maps from finite data.

To complementarily quantify the goodness of predictions in time-series forecasting tasks for Fig. \ref{fig:4}, we used the Furthest Predicted Point (FPP) \cite{Morales2021}, and the standard Root-mean square error (RMSE): For the multi-step forecasting task, let
\begin{equation}
u_{0:T-1}=(u_0,\ldots,u_{T-1})\in\mathbb V^*_{T}
\end{equation}
be an input sequence from the test set, and let \(x_T=\mathbf f(u_{0:T-1})\) be the corresponding reservoir state. Starting from \(x_T\), we generate an autonomous prediction by closing the loop through the trained readout \(\pi:\mathbb R^N\to\mathbb R^d\):
\begin{equation}
\widehat u_T=\pi(x_T),
\end{equation}
and, for \(k\geq 0\),
\begin{align}
& \widehat x_T=x_T, \\
&\widehat x_{T+k+1}
=
\phi\!\left(
\rho W^{\mathrm{res}}\widehat x_{T+k}
+
\varepsilon W^{\mathrm{in}}\widehat u_{T+k}
\right),
\\
&\widehat u_{T+k+1}
=
\pi(\widehat x_{T+k+1}).
\end{align}
The true continuation is denoted by
\[
u_{T+k}=g^{k+1}(u_{T-1}),
\]
where \(g\) is the discrete-time Lorenz flow map. Given a tolerance
\(\eta_{\mathrm{FPP}}>0\), the Furthest Predicted Point is defined as
the largest prediction time-step for which the forecast remains within
that tolerance:
\[
\mathrm{FPP}
=
\max\left\{
K>0:
\|\widehat u_{T+k}-u_{T+k}\|_2
\leq
\eta_{\mathrm{FPP}}, \forall k\leq K
\right\}.
\]
Thus, FPP measures the length of the time interval over which the
closed-loop reservoir prediction remains accurate. As a complementary error measure, we compute the root-mean-square error (RMSE) over a maximum length \(K_{\max}\):
\[
\mathrm{RMSE}
=
\left(
\frac{1}{K_{\max}+1}
\sum_{k=0}^{K_{\max}}
\|\widehat u_{T+k}-u_{T+k}\|_2^2
\right)^{1/2}.
\]
Both FPP and RMSE are then averaged over the test sequences.

\subsubsection{B.3. Driven regimes and practical neighborhood separation}

We summarize why the three non-marginal driven regimes---strong contraction,
strong instability/mixing, and saturation---are unfavorable for the
finite-resolution form of neighborhood separation needed for stable
interpolation. The point is not that each regime necessarily violates NSP as
an exact topological property, but rather that each one obstructs its practical,
computationally useful realization.

\paragraph*{(i) Strongly contractive driven regime ($\mathrm{MLE}\ll 0$).}
In a strongly contractive regime, the influence of remote input coordinates is
rapidly attenuated. Sequences that differ mainly far from the readout time can
therefore be mapped to nearly indistinguishable reservoir states. The
representation behaves as an effective finite-memory encoder: it retains only a
short recent window and suppresses distinctions carried by the distant past.
The limiting case \(\rho=0\) makes this collapse explicit, since the state
depends only on the last input coordinate. For small recurrence, the same
effect appears perturbatively, because low-order terms in \(\rho\) transmit
information only from a finite recent window.

\paragraph*{(ii) Strongly unstable or mixing driven regime.}
The opposite regime is not characterized by insufficient separation, but by
uncontrolled separation. Strong instability can amplify small input differences,
but nearby input neighborhoods may then be stretched, folded, or mixed into
highly distorted subsets of reservoir space. As a result, nearby sequences need
not remain coherent in state space, and close reservoir states may originate
from unrelated input sequences. This destroys the local geometry required for
reliable low-complexity interpolation.

\paragraph*{(iii) Saturation regime.}
For large effective gain, many state components approach the boundary of the
activation range. Input perturbations are then filtered by the derivative of the
activation function, which for \(\phi=\tanh\) becomes exponentially small in
saturated directions. Thus, the components that dominate the state vector are
also the least responsive to changes in the input. Persistent saturation
therefore produces a form of geometric collapse: distinct sequences may induce
only small state variations because input sensitivity is suppressed by the
bounded nonlinearity.

\paragraph*{(iv) Marginal driven stability.}
Near marginal driven stability, input perturbations are neither rapidly erased
nor explosively amplified. The representation varies over input neighborhoods
in a controlled but non-degenerate way: sequences remain distinguishable without
producing unstable geometry. This is the regime in which finite-resolution
proxies of NSP are expected to be both observable and useful. The response clouds
are sufficiently spread out to support interpolation, while retaining enough
local coherence for linear or low-degree polynomial readouts to generalize.

\end{document}